%% file: main.tex
\documentclass[12pt,a4paper]{article}

\input{macros}
\preprint{Imperial-TP-2025-CH-7}

\title{A New Action for Superstring Field Theory}

\author{Chris Hull}

\affiliation{Abdus Salam Centre for Theoretical Physics, The Blackett Laboratory, Imperial College London, Prince Consort Road, London, SW7 2AZ, UK}

\emailAdd{c.hull@imperial.ac.uk} 

\abstract{
Sen’s superstring field theory successfully formulates perturbative superstring theory but has a number of unusual features. The action is not  background independent and the resulting effective field theory does not appear to have standard diffeomorphism symmetry - instead there is an exotic gauge symmetry that acts on some fields but not others.  A new action is found that reduces to Sen’s action in a certain limit and gives the same results as Sen's theory for physical superstring amplitudes. However, the new theory  is background independent: there is a shift symmetry so that a change in the background can be compensated for by a corresponding shift in the string fields. Sen's theory has two string fields leading to the physical interacting superstring plus a free superstring that decouples from the physical superstring. The new terms in the action  give self-interactions to the second superstring which remains decoupled from the physical superstring. The two superstrings each have a graviton and the effective action for the two gravitons coupled to a background metric is studied and shown to be background independent and diffeomorphism invariant, and the origin of further diffeomorphism-like symmetries is explained.
}

\begin{document}

\pagestyle{myplain}
\maketitle


\section{Introduction}

String field theory (SFT) gives a  second-quantised formulation of string theory that facilitates the discussion of non-perturbative phenomena such as classical solutions and tachyon condensation. Its quantisation reproduces the correlation functions of 1st-quantised string theory while  allowing a systematic treatment of IR divergences, renormalisation, unitarity and UV finiteness. See \cite{Sen:2024nfd} for a recent review with an extensive list of references. Further reviews include \cite{Erbin:2021smf,Maccaferri:2023vns,Erler:2019loq}.
 The issue of diffeomorphism symmetry and how it is related to the symmetries of SFT together with the question of how gravity is formulated in SFT will be of particular interest here and has been addressed in \cite{Ghoshal:1991pu,Hull:2009mi,Mazel:2025fxj,Mamade:2025kfk,Mamade:2025jbs}.

The SFT for the bosonic closed string was constructed by Zwiebach \cite{Zwiebach:1992ie} and the interactions and gauge symmetry are governed by a set of $n$-brackets for integers $n\ge 1$ which satisfy an $L_\infty$ algebra. However, the generalisation of this to the superstring proved to be challenging. One obstacle was the difficulty in constructing  a kinetic term for string fields in Ramond sectors. Another was that a SFT for the type IIB superstring should give an action for the massless 4-form RR gauge field $A_4$ with  field strength $F_5=dA_4$ that is self-dual, $F_5=\ast F_5$. The construction of a covariant action for such self-dual fields has long been a challenge, yet the IIB SFT should provide such an action.
These obstacles were overcome by Sen \cite{Sen:2015uaa,Sen:2017szq} who gave  SFTs for the heterotic and type II superstrings that involved {\it two} string fields.
Having two string fields allowed the construction of a kinetic term for all sectors and it was shown that this theory gave the correct physical spectrum and perturbative correlation functions. However, having two string fields led to a doubling of the spectrum. The theory has a physical sector consisting of the interacting (heterotic or type II) superstring which includes the graviton, together with a shadow sector consisting of a free superstring that completely decouples from the physical superstring. In particular, the shadow superstring doesn't couple to the graviton that arises in the physical superstring. The SFT then gives the physical superstring, together with an extra sector that decouples from this and doesn't affect the physics. It was this doubling of the string that allowed the construction of an action.

However, the superstring field theory depends on a choice of $c=15$ superconformal field theory (SCFT) (e.g.\ the 10 free bosons $X^\mu$
 and 10 free fermions $\psi^\mu$ for for NSR formulation of the superstring in $D=10$ Minkowski space)  and whereas the  equation of motion for the physical superstring  field is independent of this choice of background \cite{Sen:2017szq}, the action does depend on the choice of background SCFT. For an SCFT defined by a non-linear sigma model,  the dependence on the SCFT  will involve a dependence on a background metric, given by the target space metric of the sigma model. 
Secondly, it was shown in \cite{Mamade:2025jbs} that the SFT gauge symmetry leads to  gauge transformations with a vector parameter that act somewhat like diffeomorphisms on some fields, but under which other fields are invariant and  yet others  transform into each other. 
These transformations were referred to as \lq exotic diffeomorphisms' in \cite{Mamade:2025jbs}.
However, {\it all} fields must transform under changes of coordinates and so all must transform under   diffeomorphisms. 
As the exotic diffeomorphisms leave some fields invariant, 
they appear to be distinct from  genuine diffeomorphisms and
the question arises as to whether the SFT has  proper diffeomorphism symmetry and so whether the theory is independent of the choice of coordinates.
This is related to fact that, in this theory, gravity is not universal in the sense that there are fields that do not couple to the physical graviton; these are the ones that are inert under the exotic diffeomorphisms. 
Moreover, the SFT gauge algebra closes on field-dependent transformations  and the same is expected to be true for the exotic diffeomorphisms (although some simplification is to be hoped for after field redefinitions) \cite{Mamade:2025jbs}.
On the other hand, general relativity is expected to be part of the low energy effective action of string theory, and the above features make it hard to see how  general relativity might emerge from this SFT.

This  superstring field theory led to 
 Sen's action \cite{Sen:2015nph,Sen:2019qit} for self-dual gauge fields, which 
 also involves a doubling of the fields.
 This theory was inspired by the SFT construction and it has recently been confirmed that it  arises directly from the SFT \cite{Mamade:2025jbs}.
This theory gives the physical gauge field $A_4$ which couples to gravity and the other physical fields, together with a second gauge field $\wt A_4$ which also has self-dual field strength but which decouples from the physical fields and in particular from $A_4$ and from gravity. Instead, $\wt A_4$ couples only to a Minkowski metric. There is a diffeomorphism-like symmetry under which the graviton transforms in the standard way but under which $\wt A_4$ is invariant,
 while
some fields have non-standard transformations in that they transform into other fields. 
These transformations are part of the \lq exotic diffeomorphisms' of \cite{Mamade:2025jbs}.

This action for self-dual gauge fields
has issues similar to those of the SFT.  Firstly, it involves the  Minkowski metric and this is highly restrictive as most spacetimes do not admit such a metric. Secondly, not all fields transform under the exotic diffeomorphisms and the theory doesn't appear to have conventional diffeomorphism symmetry.
These issues were addressed in \cite{Hull:2023dgp}, where a generalisation of Sen's action was found in which the Minkowski metric was replaced by a second metric on the spacetime. The theory has two sectors, a physical sector including the physical gauge field $A_4$ and the physical metric $g$, together with a shadow sector consisting of the second gauge field $\wt A_4$ and the second metric $\tilde g$. There are only interactions within each sector and the two sectors completely decouple from each other.
The theory has {\it two} diffeomorphism-like symmetries, one for which $g$ is the gauge field and which acts only on the physical sector and one for which
$\tilde g$ is the gauge field and which acts only on the shadow sector.
The diagonal subgroup acts on all fields and can be identified with the real diffeomorphism group that is related to coordinate independence.
The bi-metric geometry gives an understanding of 
the novel couplings that are non-linear in both metrics  \cite{Hull:2023dgp}.

In this paper, a new SFT action for the superstring will be presented which incorporates some of the features found in the bi-metric action for self-dual gauge fields. 
The new SFT again contains two decoupled sectors, but now both are superstrings with a full set of self-interactions.
As for Sen's SFT, the physical superstring and the shadow superstring are completely decoupled, so that the the shadow superstring does not affect the physics
of the physical superstring: as will be seen in section \ref{newactionsect}, the field equation for the physical superstring has no dependence on the shadow sector string field. In particular, the \lq wrong sign' for the kinetic term for the shadow sector has no consequences for the physical superstring.
The Sen SFT is recovered when the coupling constant for the self-interactions of the shadow superstring is set to zero, so that it becomes a free theory, while the physical sector remains fully interacting.
However, with interactions in both sectors, the action is independent of the choice of background   and  has two   exotic diffeomorphism symmetries, one for each sector. The diagonal subgroup acts on all fields and  is related to conventional diffeomorphism symmetry.  The construction will be motivated by an analysis of the effective actions for the gravitons that arise. General relativity achieves a background-independent formulation of a graviton field $h$ in a background spacetime 
$\bar g$ by depending only on the total metric $g=\bar g+h$ and, in a similar spirit, the SFT presented here is background independent in the sense that it can be viewed as depending only on total string fields and not separately on a background  and   fluctuations. For the action given here, both sectors are expanded about the same background, but the full theory is independent of this choice of background. This can be generalised to an action in which each sector is expanded around a different background; this is discussed in section \ref{Discussion}.
The notion of background independence used in this paper is the limited one that is used in the string field theory literature (see \cite{Sen:2017szq,Sen:1990hh,Sen:1993mh,Sen:1993kb,Sen:1994kx}) and which is reviewed in section \ref{backin}. This formulation is not as general as one might like, and this is discussed in section \ref{Discussion}.

\section{Sen's Type II  Superstring Field Theory}

Classical type II superstring  field theory is constructed from a (1,1) SCFT with $c=15$, coupled to the usual $b,c$ ghosts and the 
fermions
$(\xi,\eta)$ and  scalar $\phi$ obtained by bosonising the superghosts
 $\beta,\gamma$ to give a  SCFT with total central charge $c=0$.
 The string field is a state in the Hilbert space of the SCFT. 
The string fields are then acted on by the BRST charge $Q$ and the super-Virasoro generators of the 2-d SCFT.

Sen's type II superstring field theory	uses two  
	string fields, $\Psi$ and $\widetilde\Psi$.
Picture number $-1$ is used for  both the left- and right-moving  NS sectors,   picture number $-1/2$ is used for the left- and right-moving  R sectors of the 
$\Psi$ field, while  picture number $-3/2$ is used for the left- and right-moving  R sectors of the 
$\widetilde \Psi$ field.
Denoting 
the space of string states of left-moving picture number $p$  and 
right-moving picture number $q$ by ${\cal H}_{p,q}$, the
string fields take values in the following spaces:
	\be
	\label{Hilbs}
	\begin{split} 
		& \Psi\in {\cal H}_c \equiv {\cal H}_{-1,-1}\oplus {\cal H}_{-1,-{\frac 1 2 }}\oplus {\cal H}_{-{\frac 1 2},-1}\oplus {\cal H}_{-{\frac 1 2},-{\frac 1 2}}, \\[0.6ex] 
		& \widetilde\Psi\in \widetilde{\cal H}_c \equiv
		{\cal H}_{-1,-1}\oplus {\cal H}_{-1,-{\frac 3 2}}\oplus {\cal H}_{-{\frac 3 2},-1}\oplus {\cal H}_{-{\frac 3 2},-{\frac 3 2}}\, .
	\end{split}
	\ee
	For the classical SFT, both string fields $\Psi$ and $\widetilde\Psi$ are Grassmann even, have ghost number two and   are annihilated by 
	\be 
	\label{Virg}
	L_0^-= L_0- \bar L_0,\qquad b_0^- =b_0 - \bar b_0\, .
	\ee
	They are also required to be even under the action of the IIA or IIB GSO operator (see \cite{Mamade:2025jbs} for the definitions of these operators), giving the IIA or IIB superstring respectively.
	The fields then take value in the subspace ${\cal H}$ of ${\cal H}_c\oplus \widetilde{\cal H}_c$ defined by these constraints.

SCFT correlators on the 2-sphere can be used to define a map ${\cal H}_c^n\to \widetilde{\cal H}_c$ for each integer $n\ge 2$, denoted by an $n$-bracket $[A_1,A_2,\dots A_n]$. 
See \cite{Sen:2015uaa,Sen:2017szq} for the definition and properties of these brackets. This infinite set of brackets gives what has been called a twisted $L_\infty$ algebra in \cite{Singh:2024mek}.
A SFT inner product $\langle A \,, B \rangle$ is defined by the BPZ inner product on the SCFT with   an insertion of $c_0^- =c_0 - \bar c_0$, and this can be used to define a map 
${\cal H}_c^n\to \mathbb{C}$
\be\label{esquare}
\{A_0,\cdots, A_n\}=\langle A_0\,, \, [A_1,\cdots ,A_n]\, \rangle  \,.
\ee
Following \cite{Sen:2015uaa,Sen:2017szq,Mamade:2025jbs}, the $n$-bracket $[A,\dots,A]$ will be denoted $[A^n]$ and $\langle A\,, \, [A^{n-1}]\, \rangle$ will be denoted $\{A^n\}$.
	
Sen's classical action is  
\be\label{senact}
	S=- \tfrac{1}{2}  \langle \widetilde \Psi ,  
	Q \, {\cal G} \, \widetilde\Psi\rangle  
	+ \langle \widetilde \Psi ,   
	Q \, \Psi\rangle +\sum_{n=3}^\infty 
	 {\frac {\kappa^{n-2}}  {n!}}
	 \{\Psi^n\}\, ,
	\ee
	where $\kappa$ is a coupling constant.
	  The operator ${\cal G}:  \widetilde{\cal H}_c \to {\cal H}_c$  changes the picture number  and is defined by:
	\be
	\label{Gdefpc}
	{\cal G}\equiv \begin{cases} \hbox{${\bf 1}\ $   on ${\cal H}_{-1,-1}$}\, ,\cr \hbox{${\cal X}_0$ on ${\cal H}_{-1,-3/2}$}\, ,
		\cr \hbox{$\bar{\cal X}_0$ on ${\cal H}_{-3/2,-1}$}\, ,\cr \hbox{${\cal X}_0\bar{\cal X}_0$ on ${\cal H}_{-3/2,-3/2}$.}
	\end{cases}
	\ee
	where
	the
	 zero modes of the left-moving and right-moving picture changing operators are
	${\cal X}_0$ and $\bar{\cal X} _0$.
	The  string fields satisfy the reality conditions given in \cite{Sen:2016bwe} which ensure that the action is real.
	
	The gauge symmetry of the classical theory is
	\be
	\label{gaugetrans}
	\begin{split}
		\delta   |\widetilde\Psi\rangle  =& \  Q |\widetilde \Lambda\rangle  + \sum_{n=1}^\infty  {\kappa ^{n-1}\over n!}  \, [ \Lambda \Psi^n ]\,, \\ 
		\delta  |\Psi\rangle  =& \  Q | \Lambda\rangle  + \sum_{n=1}^\infty  {\kappa ^{n-1}\over n!}  \, {\cal G} \,  [ \Lambda \Psi^n ]
		\, ,
	\end{split}
	\ee
	with Grassmann odd gauge parameters $(\Lambda, \widetilde\Lambda) \in ({\cal H}_c, \widetilde{\cal H}_c)$ of ghost number one.
	The gauge algebra  was analysed in \cite{Mamade:2025kfk}, where it was found that the commutator of two SFT gauge transformations gives SFT gauge transformations with string-field parameters   that depend on the string fields, together with an on-shell-trivial symmetry, so that the algebra closes on-shell with field-dependent structure functions.
	
	Varying  the 
action~(\ref{senact})  gives
the field equations
 \be
    Q \Psi - Q {\cal G} \widetilde \Psi=0\,,
    \ee
    \be
   Q \widetilde \Psi + \sum_{n=2}^\infty {\kappa ^{n-1}\over n!} [ \Psi^n] =0\, ,
   \label{tipseq}
 \ee
 which then imply
  \be
  \label{psieq}
 Q \Psi +{\cal G}\sum_{n=2}^\infty {\kappa ^{n-1}\over n!} [ \Psi^n] =0 
 \ee
and
 \be
 \label{freeeq}
Q\wh    \Psi=0
\ee
where
\be
\wh    \Psi= \Psi- {\cal G} \widetilde \Psi\, .
  \ee
The gauge transformations (\ref{gaugetrans}) then give
\be \delta   |\wh  \Psi\rangle  = \  Q |\wh   \Lambda\rangle
\ee
where
\be
\wh   \Lambda=  \Lambda-{\cal G} \widetilde\Lambda
\ee
and this is, of course, a symmetry of the field equation (\ref{freeeq}).

A particular set of solutions that were considered in \cite{Sen:2017szq} take $\Psi$ to be a solution of (\ref{psieq}) and then take $\wt\Psi=\Psi$. This requires that $\Psi$
is in the NS-NS sector and implies $\wh \Psi=0$.

Setting $\kappa=0$ gives a free theory with quadratic action
\be\label{free}
	S_{free}=- \tfrac{1}{2}  \langle \widetilde \Psi ,  
	Q \, {\cal G} \, \widetilde\Psi\rangle  
	+ \langle \widetilde \Psi ,   
	Q \, \Psi\rangle \, ,
	\ee
with symmetries
\be
	\label{gaugetransfree}
	\begin{split}
		\delta   |\widetilde\Psi\rangle  =& \  Q |\widetilde \Lambda\rangle
		  \,, \\ 
		  \delta   |\widehat\Psi\rangle  =& \  Q |\widehat \Lambda\rangle
		  \,, \\ 
		\delta  |\Psi\rangle  =& \  Q | \Lambda\rangle
		\,. 
	\end{split}
	\ee

Note that the field equations (\ref{psieq}),(\ref{freeeq}) for $\Psi$ and $\widehat\Psi$  are completely decoupled. Then  $\Psi$ is the string field for the physical interacting type II superstring  
while $\widehat \Psi$ is the string field for a second type II superstring which is a free theory.
The spectra of the $\Psi$ and $\widehat \Psi$ strings are identical but one is an interacting theory and the other is a free theory.
In the NS-NS sector ${\cal G}=1$ and $\Psi$ and $\widehat\Psi$ can be regarded as the independent fields, with $\widetilde \Psi=\Psi-\widehat\Psi$.
In the other sectors, given  $\Psi$ 
satisfying (\ref{psieq}), $\widetilde \Psi$ is found by solving (\ref{tipseq}) and then  $\widehat\Psi$ will satisfy (\ref{freeeq}).
In the RR sector, $\Psi$ and $\widehat\Psi$ give field strengths for the massless RR fields and $\widetilde \Psi$ gives  gauge potentials for  linear combinations of these field strengths (see  \cite{Mamade:2025jbs}).
Then (in all but the NS-NS sector) $\Psi$ and $\widetilde \Psi$  are the independent fields and $\widehat\Psi$ is a field derived from these.

This construction  of the string field theory  depends on a choice of SCFT background and an important issue is whether the final theory has some independence of the choice of background.

\section{Free Graviton Action}

For the superstring in $D=10$ Minkowski space, the matter SCFT consists of 10  scalar fields $X^\mu$ and $10 $  fermionic fields $\psi^\mu$.
In \cite{Mamade:2025jbs}, the SFT action was  partially evaluated for massless bosonic  spacetime  fields in the NS-NS and RR sectors  to lowest order in $\alpha '$ and to cubic order in fields, giving the bosonic sector of the type II supergravity theory to first order in $\kappa$.\footnote{In \cite{Mamade:2025jbs}, 
$\wt\Psi $ was set equal to $\Psi$ in the NS-NS sector, and the NS-NS and RR kinetic terms were then evaluated, together with cubic terms with one NS-NS field and two RR fields.} The calculation of  \cite{Mamade:2025jbs} follows closely the corresponding calculation for the bosonic string in \cite{Hull:2009mi}, and a key feature is that the physical fields mix with auxiliary fields and it is only when the auxiliary fields are eliminated that the standard action is obtained.
Here we will focus only on the terms involving the graviton.

The graviton $h_{\mu\nu}$ appears through the string field
\be
		\label{gravv}
					\Psi_{grav}=  \int\frac{d^Dp}{(2\pi)^D}
			\biggl(\, 
			\tfrac{1}{2}\, 
			h_{\mu\nu}(p)\ c\Bar{c}\, \psi^\mu\Tilde{\psi}^\nu e^{-\phi}e^{-\bar \phi} 
			 \biggr)e^{ip\cdot X}\,. 
		\ee
In \cite{Mamade:2025jbs}, the NS-NS sector was restricted  to the subspace in which $\Psi=\widetilde \Psi $.
Then setting $\Psi=\widetilde \Psi=  \Psi_{grav}+\dots $ (where the $\dots$ includes couplings to certain auxiliary fields) in the free action (\ref{free})
led to the linearised Einstein action
\be
		\label{gravact}
		S_{lin}[h,\eta]= \tfrac{1}{2}  \int d^{10}x
		\left(   h^{\mu\nu}\partial^2 h_{\mu\nu}+2(\partial^\nu h_{\mu\nu})^2  -   \, h \partial^2 h  
+ 2 h \,\partial_\mu \partial_\nu \, h^{\mu\nu}
		  \right)     \,
		\ee	 
		where $h=\eta ^{\mu\nu}h_{\mu\nu}$.
The SFT gauge invariance	gives rise to	the linearised diffeomorphism symmetry
\be
		\delta h_{\mu\nu}  = \partial_\mu \lambda_\nu + \partial_\nu \lambda_\mu \,.
		\ee

As was to be expected, the action for the graviton to this order
agrees with the expansion of the Einstein action
\be
\label{EH}
S_{grav}= \frac{1}{2 \kappa ^2}  \int d^{10}x \sqrt{-g}R
\ee
for metric 
\be
g_{\mu\nu}=\eta_{\mu\nu}+\kappa h_{\mu\nu}
\ee
to quadratic order in the graviton field. 
In  \cite{Hull:2009mi}, the calculation for the bosonic string was extended to cubic terms arising from the terms of order $\kappa$ in the  string field theory action and these terms were found to be in agreement with the expansion of (\ref{EH}) to linear order in $\kappa$. 

Consider now the full linearised theory (\ref{free}) in which one does not restrict to the sector in which $\Psi=\widetilde \Psi $.
Then there is another symmetric tensor gauge field $\tilde h_{\mu\nu}$ arising from $\widetilde \Psi $ with
\be
		\label{gravvt}
					\widetilde\Psi_{grav}=  \int\frac{d^Dp}{(2\pi)^D}
			\biggl(\, 
			\tfrac{1}{2}\, 
			\tilde h_{\mu\nu}(p)\ c\Bar{c}\, \psi^\mu\Tilde{\psi}^\nu e^{-\phi}e^{-\bar \phi} 
			 \biggr)e^{ip\cdot X}\,. 
		\ee
The calculation of \cite{Mamade:2025jbs} can be extended to include both $h$ and $\tilde h$. The result for the  action for the two gravitons is
\be
\label{gravact2}
S_{grav}=
S_{lin}[h,\eta]-S_{lin}[\hat h,\eta]
 \ee	 		
where 
\be
\hat h_{\mu\nu} =h_{\mu\nu}-\tilde h_{\mu\nu}
\ee		
so that  $\hat h$ can be viewed as a graviton mode for the string field $\wh    \Psi= \Psi- {\cal G} \widetilde \Psi$. Here $S_{lin}$ is defined by (\ref{gravact}) and $S_{lin}[\hat h,\eta]$ is given by replacing $h$ with $\hat h$ in  (\ref{gravact})  with $\hat h$.
It was seen above that the string fields $\Psi,\wh    \Psi$ satisfy decoupled equations of motion which, to linear order, are simply
$Q\Psi=0$, $Q\wh    \Psi=0$, so that it was to be expected that there should be two gravitons, one from $\Psi$ and one from $\wh    \Psi$.
The field equation for both $h_{\mu\nu}$ and $\hat h_{\mu\nu}$ (and also for $\tilde h_{\mu\nu}$) is the linearised Einstein equation.
However, the action for the  graviton $h$ has the correct  sign for a unitary theory  while that for $\hat h$ has the opposite sign.
This action has two gauge invariances, one with parameter $\lambda_\mu$ for which $h$ is the gauge field and one
with parameter $\hat \lambda_\mu$
 for which  $\hat h$ is the gauge field:   
\be
\delta h_{\mu\nu}  = \partial_\mu \lambda _\nu + \partial_\nu \lambda_\mu
\qquad 
\delta \hat h_{\mu\nu}  = \partial_\mu \hat \lambda_\nu + \partial_\nu \hat \lambda_\mu
 \,.
		\ee 
As the notation suggests, the $\lambda_\mu$ symmetry arises from the  SFT symmetry (\ref{gaugetransfree})  with parameter $\Lambda$ 
while the $\hat \lambda_\mu$ symmetry arises from the  SFT symmetry (\ref{gaugetransfree})   with parameter $\hat \Lambda$.

\section{Gravitons, Symmetry and Background Independence}

\subsection{Symmetry and Gravity from String Theory}

The bosonic string field theory action gives an action for the graviton that agrees with the Einstein-Hilbert action to cubic order in the graviton, after eliminating auxiliary fields and performing some field redefinitions, while the  string field theory gauge symmetry gives  a transformation that agrees with the expected diffeomorphism symmetry to linear order in the graviton \cite{Hull:2009mi}.
The results of \cite{Mamade:2025kfk} show that the graviton gauge symmetries that arise from the SFT satisfy the diffeomorphism algebra to lowest order, but the algebra gives field-dependent transformations combined with on-shell trivial symmetries to higher order. It is not yet known whether further field redefinitions can relate the SFT symmetries to diffeomorphisms, or whether the algebra reflects a more exotic action that differs from the Einstein-Hilbert  action at higher orders; see \cite{Mamade:2025kfk,Mamade:2025jbs} for further discussion.
This means that it is difficult at present to talk about the gravity action from SFT beyond lowest order.
Nonetheless, there is good reason to believe that the Einstein-Hilbert action arises as part of the low-energy effective action for string theory.
This is supported by calculations in the 1st-quantised formulation, which gives the Einstein-Hilbert  action plus a series of higher-derivative corrections arising in an $\alpha '$ expansion.

In this section, the graviton actions arising from superstring field theory will be considered. The graviton action arising from the SFT   is known to lowest order in the graviton fields. However, the effective action for the physical graviton is again the Einstein-Hilbert  action (plus   higher-derivative corrections) and this will enable us to discuss the effective actions for the full superstring field theory. This effective action is expected to arise from the SFT after integrating out all other fields, and it would be interesting to verify this in detail.

\subsection{Gravitons in a Background Spacetime}

Suppose now that the matter $c=15$ SCFT is given by a non-linear sigma model with $D$-dimensional target space $M$ with metric $\bar g _{\mu\nu}(X)$.
In the simplest case, which will be considered here, $D=10$ and the metric $\bar g _{\mu\nu}(X)$ is Ricci-flat.\footnote{More generally, other background fields (such as a $B$-field and dilaton) can be included, the metric satisfies an Einstein equation with sources from the other fields and with higher derivative corrections,  and the dimension $D$ need not be $10$.}
The free SFT action (\ref{free}) 
 for the sigma model SCFT should give a quadratic action for fields coupled to the background metric $\bar g _{\mu\nu}(X)$.
The action (\ref{gravact2}) for two free gravitons propagating in Minkowski space should then be replaced by a quadratic action  for two  free  gravitons propagating in $M$.
The natural 2-derivative action  that reduces to (\ref{gravact}) in Minkowski space is
\be
\label{gravact2cov}
S_{grav}= S_{lin}[h,\bar g]-S_{lin}[\hat  h,\bar g]
\ee	
where
\be
		\label{gravactX}
		S_{lin}[h,\bar g]= \tfrac{1}{2}  \int d^{10}x \sqrt {-\bar g}
		\left(   h^{\mu\nu}\bar\nabla^2 h_{\mu\nu}+2(\bar\nabla^\nu h_{\mu\nu})^2  -   \, h\bar \nabla^2 h  
+ 2 h \,\bar\nabla _\mu \bar\nabla_\nu \, h^{\mu\nu}
		  \right)     \,
		\ee	 
Here indices are raised, lowered and contracted using the background metric $\bar g _{\mu\nu}$, while $h=\bar g ^{\mu\nu}h_{\mu\nu}$ and $\bar \nabla$ is the Levi-Civita connection for $\bar g _{\mu\nu}$.  
The form of the first term $S_{lin}[h,\bar g]$ follows from graviton correlation functions in the 1st quantised approach, while symmetry between the two sectors and the requirement 
that the action reduces to (\ref{gravact}) in Minkowski space 
then determines the second term $S_{lin}[\hat h,\bar g]$.

The sigma-model is invariant under diffeomorphisms of the target space and this is reflected in  the diffeomorphism symmetry of the action
\be
\delta \bar g _{\mu\nu}= {\cal L}_\xi  \bar g _{\mu\nu} = \bar \nabla_\mu \xi _\nu + \bar \nabla_\nu \xi_\mu, \qquad \delta  h _{\mu\nu}= {\cal L}_\xi   h_{\mu\nu},
\qquad \delta  \hat h _{\mu\nu}= {\cal L}_\xi   \hat h_{\mu\nu}
\ee
with $ \bar g _{\mu\nu},h _{\mu\nu},\hat h _{\mu\nu}$ all transforming tensorially. Here ${\cal L}_\xi$ the Lie derivative with respect to an infinitesimal vector field $\xi$.
If $ \bar g _{\mu\nu} $ satisfies the Einstein equation, then the action (\ref{gravact2cov}) is also  invariant under  
\be
\delta h_{\mu\nu}  = \bar \nabla_\mu \lambda _\nu + \bar \nabla_\nu \lambda_\mu
\qquad 
\delta \hat h_{\mu\nu}  = \bar \nabla_\mu \hat \lambda_\nu + \bar \nabla_\nu \hat \lambda_\mu
 \,.
		\ee 
Thus the theory is diffeomorphism invariant, but also has  a gauge invariance for each graviton.
The theory is clearly background dependent: the theory depends on the choice of $\bar g _{\mu\nu}$.

\subsection{Background Independence for Einstein Gravity}

The free graviton action $S_{lin}[h,\bar g]$ given by (\ref{gravactX})
 can be promoted to a background-independent action
  by adding interactions.
 The Einstein-Hilbert action
 \be
 \label{SEH}
 S_{EH}(g)= \frac 1 {2\kappa^2}\int d^{10}x \sqrt {-g} R(g)
 \ee
 with
 \be
g _{\mu\nu}= \bar g _{\mu\nu}+\kappa h_{\mu\nu}
\ee
gives (\ref{gravactX}) plus terms of all orders in $\kappa$ and depends only on the total metric $g _{\mu\nu}= \bar g _{\mu\nu}+\kappa h_{\mu\nu}$. It has the diffeomorphism symmetry
\be
\label{diffy}
\delta   g _{\mu\nu}= {\cal L}_\xi    g _{\mu\nu} =   \nabla_\mu \xi _\nu +   \nabla_\nu \xi_\mu 
\ee
together with the shift symmetry
\be
\label{shift}
\delta h_{\mu\nu}= \alpha _{\mu\nu},\qquad
\delta \bar g_{\mu\nu}= -\kappa\alpha _{\mu\nu}
\ee
The transformation (\ref{diffy}) can be taken to act geometrically on the background and fluctuation:
\be
\label{diffy2}
\delta   \bar g _{\mu\nu}= {\cal L}_\xi  \bar  g _{\mu\nu} =  \bar \nabla_\mu \xi _\nu +  \bar \nabla_\nu \xi_\mu , \qquad
\delta   h_{\mu\nu}= {\cal L}_\xi    h _{\mu\nu} 
\ee
so that they both transform as tensors.
Then
choosing $\xi_\mu =\lambda _\mu$, $\alpha _{\mu\nu}= \kappa^{-1}{\cal L}_\lambda \bar g_{\mu\nu}$ gives
\be
\label{gaugesss}
\delta h_{\mu\nu}=  \nabla_\mu \lambda _\nu +   \nabla_\nu \lambda _\mu ,\qquad
\delta \bar g_{\mu\nu}= 0
\ee
Thus the background independence allows us to relate the gauge symmetry (\ref{gaugesss}) to the diffeomorphism symmetry (\ref{diffy}).

\subsection{Background Independence for  the Physical Sector SFT Gravitons }

The string action (\ref{senact}) contains terms   $\kappa ^{n-2}\Psi ^n$ of all orders in the string field for all $n \ge 2$ and so should give rise to a graviton action contain terms    $\kappa ^{n-2}h ^n$
of all orders in the graviton  field $h$. It was shown in \cite{Mamade:2025jbs} (following a similar calculation for the bosonic string in \cite{Hull:2009mi}) the term of order $  h ^2$ agrees with the action (\ref{SEH}) expanded to zero'th order in $\kappa$. It is thus to be expected that the action (\ref{SEH})  should be part of an effective field theory for the string.
However, the SFT action is only of quadratic order in $\tilde \Psi$ and this results in an action quadratic in $\hat h$.
The natural action that is non-polynomial in $h$  but quadratic in $\hat h$   in a background  with metric $\bar g$
is
\be
\label{2gra1}
S=S_{EH}(\bar g  +\kappa h )- 
S_{lin}[\hat  h,\bar g]
 \ee
 which is non-linear in $h$ but quadratic in  $\hat h$ and $\tilde h$.
 The Einstein-Hilbert action is background independent, depending only on the total metric $g=\bar g  +\kappa h$, but the quadratic term depends explicitly on the background metric $\bar g$.
 The action is invariant under diffeomorphisms
 \be
 \label{difff}
\delta \bar g _{\mu\nu}= {\cal L}_\xi  \bar g _{\mu\nu} , \qquad \delta  h _{\mu\nu}= {\cal L}_\xi   h_{\mu\nu},
\qquad \delta  \hat h _{\mu\nu}= {\cal L}_\xi   \hat h_{\mu\nu}
\ee
so that 
\be
\delta   g _{\mu\nu}= {\cal L}_\xi    g _{\mu\nu} =   \nabla_\mu \xi _\nu +   \nabla_\nu \xi_\mu 
\ee
and these are inherited from the diffeomorphism invariance of the CFT.
In addition, there is the 
symmetry
\be
\label{exo1}
\delta h_{\mu\nu}=  \nabla_\mu \lambda _\nu +   \nabla_\nu \lambda _\mu ,\qquad
\delta \bar g_{\mu\nu}= 0, \qquad \delta \hat h_{\mu\nu}  =0
\ee
corresponding to the SFT symmetry (\ref{gaugetrans}) with parameter $\Lambda$ and,
if  the background metric $\bar g$ is Ricci-flat, there is
a symmetry 
 \be
 \label{exo2}
\delta h_{\mu\nu}  = 0
\qquad 
\delta \hat h_{\mu\nu}  = \bar \nabla_\mu \hat \lambda_\nu + \bar \nabla_\nu \hat \lambda_\mu,\qquad
\delta \bar g_{\mu\nu}= 0
		\ee 
corresponding to the SFT symmetry (\ref{gaugetrans})  with parameter $\wh   \Lambda$.

This theory is not background independent -- the kinetic term for $\hat h$ depends explicitly on the background $
\bar g$.
As a result, the gauge symmetry (\ref{exo1}) can no longer be related to the  diffeomorphism symmetry.
The field $h$ has the standard transformation for a graviton, but the field $\hat h$ is invariant. 

In  \cite{Mamade:2025jbs}, the SFT was used to find
the action   and 
  gauge transformations for the massless RR fields to first order in $\kappa$ in a background Minkowski space, $\bar g = \eta$.
  The fields appearing in the action (denoted $Q,P$ in \cite{Mamade:2025jbs}) have unusual transformations under the $ \lambda, \bar  \lambda$ symmetries (with the variation of $P$ depending on $Q$, for example). 
 The $ \lambda, \bar  \lambda$ transformations, which act like diffeomorphisms on some fields, leave other fields invariant and transform RR fields into each other. These
 are the exotic diffeomorphisms of \cite{Mamade:2025jbs}.
  If the Minkowski background is replaced by an on-shell background with metric $\bar g$, the action and gauge transformations of the RR fields is of the form discussed in \cite{Hull:2023dgp}. 
  
  Here we have seen how the exotic diffeomorphism symmetries (\ref{exo1}),(\ref{exo2}) arise  and that there is also a conventional diffeomorphism symmetry (\ref{difff}).
  For the gravity action (\ref{SEH}), the background independence allowed the exotic symmetry (\ref{gaugesss}) to be seen as a conventional diffeomorphism  combined with a shift transformation symmetry.
  However, here the background dependence implies that there is no shift symmetry and
  so the exotic symmetries for the 2-graviton theory cannot be understood in this way.
  However, in the next subsection it will be seen that modifying the action to obtain background independence does allow these symmetries to be related to diffeomorphisms.

\subsection{Background Independence for the   SFT Gravitons }

There is a natural way to regain background independence and hence a relation between the gauge symmetries and diffeomorphisms.
The quadratic action  $ S_{lin}[\hat  h,\bar g]$ for $\hat h$ can be made background independent by adding non-linear terms in $\hat h$ to construct the Einstein Hilbert action
\be
S_{EH}(\hat g)= \frac 1 {2 \hat \kappa^2}\int d^{10}x \sqrt {-\hat g} R(\hat g)
 \ee
 depending only on the total metric  
 \be
\hat g _{\mu\nu}= \bar g _{\mu\nu}+\hat\kappa \hat h_{\mu\nu}
\ee
where a new coupling constant $\hat\kappa$ has been introduced. This action
gives the quadratic action plus terms of order $\hat\kappa$ and it has the diffeomorphism symmetry
\be
\delta  \hat  g _{\mu\nu}= {\cal L}_\xi  \hat   g _{\mu\nu}  
\ee
together with the shift symmetry
\be
\delta \hat h_{\mu\nu}=\hat  \alpha _{\mu\nu},\qquad
\delta \bar g_{\mu\nu}= -\hat \kappa\hat \alpha _{\mu\nu} \, .
\ee
Then, as before, these combine to give the symmetry
\be
\delta \hat h_{\mu\nu}=  \hat \nabla_\mu\hat  \lambda _\nu +  \hat  \nabla_\nu \hat \lambda _\mu ,\qquad
\delta \bar g_{\mu\nu}= 0 \, .
\ee
where $\hat \nabla $ is the Levi-Civita connection for $\hat  g $.

The resulting  action for the two gravitons  is then the difference between two Einstein-Hilbert actions
\be
\label{2grav}
S
=S_{EH}(g )-  S_{EH}(  \hat g )
 \ee
 depending only on the total metrics 
 \be
   g _{\mu\nu}= \bar g _{\mu\nu}+ \kappa   h_{\mu\nu}
 ,\qquad
\hat g _{\mu\nu}= \bar g _{\mu\nu}+\hat\kappa \hat h_{\mu\nu}
\ee
There are two coupling constants $\kappa $ and $\hat \kappa$.
Such actions were discussed in \cite{Hull:2023dgp} and have interesting symmetry properties.
 It  is invariant under the diffeomorphisms
  \be
  \label{rdif}
 \delta  g _{\mu\nu}= {\cal L}_\xi   g_{\mu\nu},
\qquad \delta  \hat g _{\mu\nu}= {\cal L}_\xi   \hat g_{\mu\nu}
\ee
and the shift symmetry
\be
\label{shift2}
\delta \bar g_{\mu\nu}= - \alpha _{\mu\nu}
,\qquad
\delta  h_{\mu\nu}=\frac 1 \kappa  \alpha _{\mu\nu}
,\qquad
\delta \hat h_{\mu\nu}=\frac 1 {\hat \kappa}   \alpha _{\mu\nu}
\ee
reflecting independence of the background.
As the action is the sum of two decoupled terms, it is also   invariant under the $\rho$-diffeomorphisms that act on $g$ (and all fields that couple to $g$)
with
 \be \label{dff1}
 \delta  g _{\mu\nu}= {\cal L}_\rho  g_{\mu\nu},
\qquad \delta  \hat g _{\mu\nu}= 0
\ee
together with the $\hat \rho$-diffeomorphisms that act on $\hat g$ (and all fields that couple to $\hat g$) with
 \be  \label{dff2}
 \delta  g _{\mu\nu}= 0,
\qquad \delta  \hat g _{\mu\nu}= {\cal L}_{\hat \rho}   \hat g_{\mu\nu}
\ee
Setting $\kappa=\hat \kappa$, the  actual diffeomorphisms (which can be interpreted as   coordinate transformations) arise as the diagonal subgroup   in which $\xi=\rho=\hat \rho$.

The transformation (\ref{dff1}) can be realised as a transformation acting only on $h$, giving the symmetry
\be
\label{lams}
\delta h_{\mu\nu}=  \nabla_\mu \lambda _\nu +   \nabla_\nu \lambda _\mu ,\qquad
\delta \hat h_{\mu\nu}= 0,\qquad
\delta \bar g_{\mu\nu}= 0
\ee
while the transformation (\ref{dff1}) can be realised as a transformation acting only on $\hat h$, giving the symmetry
\be
\label{lamsh}
\delta \hat h_{\mu\nu}=  \hat \nabla_\mu\hat  \lambda _\nu +  \hat  \nabla_\nu \hat \lambda _\mu ,\qquad
\delta h_{\mu\nu}=0,\qquad
\delta \bar g_{\mu\nu}= 0\, .
\ee
These   exotic diffeomorphisms now have a clear origin from the $\rho$-diffeomorphisms and $\hat \rho$-diffeomorphisms  (\ref{dff1}),(\ref{dff2}). With $\kappa=\hat \kappa$, the diagonal subgroup in which $\lambda=\hat \lambda$ is a diffeomorphism (\ref{rdif}) combined with a shift symmetry (\ref{shift2}), so the background independence means that the diagonal subgroup can be identified with the diffeomorphism symmetry.

Combining (\ref{lams}) with a shift transformation (\ref{shift2}) gives
\be
 \label{difffty}
\delta \bar g _{\mu\nu}= {\cal L}_\lambda  \bar g _{\mu\nu} = \bar \nabla_\mu \lambda _\nu +  \bar \nabla_\nu \lambda_\mu, \qquad \delta  h _{\mu\nu}= {\cal L}_\lambda   h_{\mu\nu},
\qquad \delta  \hat h _{\mu\nu}= -\frac 1 {\hat \kappa} {\cal L}_\lambda  \bar g _{\mu\nu}
\ee
under which $\bar g,h$ transform tensorially but $\hat h$ does not. These are the analogue of the transformations (\ref{diffy2}).

More generally,   the background independent action $S(g,\hat g)$ depending on the two metrics $g$ and $\hat g$ could be expanded around two different backgrounds $\bar g, \bar {\hat g}$ as 
\be
\label{genbac}
   g _{\mu\nu}= \bar g _{\mu\nu}+ \kappa   h_{\mu\nu}
 ,\qquad
\hat g _{\mu\nu}= \bar {\hat g} _{\mu\nu}+\hat\kappa \hat h_{\mu\nu}
\ee
giving rise to an action $S(\bar g, \bar {\hat g},h, \hat h)$ with two shift symmetries
\be
\label{shift3}
\delta \bar g_{\mu\nu}= - \alpha _{\mu\nu}
,\qquad
\delta  h_{\mu\nu}=\frac 1 \kappa  \alpha _{\mu\nu}
\ee
and
\be
\label{shift4}
\delta \bar {\hat g}_{\mu\nu}= -\hat \alpha _{\mu\nu}
,\qquad
\delta  \hat h_{\mu\nu}=\frac 1 \kappa \hat \alpha _{\mu\nu}
\ee
There are then $\lambda $ and $\hat\lambda $ symmetries given by (\ref{lams}),(\ref{lamsh}) together with $\delta \bar {\hat g}=0$.
The $\lambda $ symmetry (\ref{lams}) is a $\rho$-diffeomorphism (\ref{dff1}) combined with a $\alpha$-shift (\ref{shift3}), while the $\hat\lambda $ symmetry (\ref{lamsh})  is a $\hat\rho$-diffeomorphism (\ref{dff2}) combined with a $\hat\alpha$-shift (\ref{shift4}).
The independent symmetries can then be taken to be the diffeomorphisms (\ref{dff1}),(\ref{dff2}) acting on the total fields, together with the two shift symmetries (\ref{shift3}),(\ref{shift4}) that arise when expanding around a background.

\section{New SFT Action for the Superstring} \label{newactionsect}

\subsection{Background Dependence and Interactions}

Sen's SFT  for the superstring is an interacting theory for $\Psi$ but a free theory for the other string field.
It was shown in \cite{Sen:2017szq} that the field equation for $\Psi$ is background independent,  but that both  the field equation for $\widetilde\Psi$ and the action are background dependent.
This can be understood from the effective action (\ref{2gra1}) that Sen's theory is expected to give for the two gravitons. The action for $h$ is background independent as it depends only on $g=\bar g +\kappa h$, while that for $\hat h$ is background dependent  as it depends explicitly on the background metric $\bar g$.
For the theory of two gravitons, background independence was obtained by  replacing (\ref{2gra1}) with (\ref{2grav}), adding interactions for the second graviton $\hat h$.

The free SFT (\ref{free}) for a particular background CFT corresponds to the free graviton action (\ref{gravact2cov}) depending on the background metric $\bar g$.
Adding the interaction term  \be\sum_{n=3}^\infty 
	 {{\kappa^{n-2}} \over {n!}}
	 \{\Psi^n\}\ee to the free action to obtain the Sen action (\ref{senact}) attained background independence for the field equation
	 \be
	 \label{psieqm}
 Q \Psi +{\cal G}\sum_{n=2}^\infty {\kappa ^{n-1}\over n!} [ \Psi^n] =0 
 \ee
  of the physical string field $\Psi$
	 and this corresponds to adding the Einstein-Hilbert interactions $\sum {\kappa^{n-2}} h^n$   for $h$ to give the action (\ref{2gra1}).
	 For the two-graviton action, full background independence was obtained by adding also interactions $\sum {\hat \kappa^{n-2}} \hat h^n$   for $\hat h$
	 to obtain the action (\ref{2grav}).
	 For the superstring SFT, this suggests adding couplings so that the field equation for $\wh \Psi$
becomes 
  \be 
  Q   \wh   \Psi + {\cal G}\sum_{n=2}^\infty 
	 {{\hat\kappa^{n-1}} \over {n!}}
	 [\wh   \Psi^n]=0\,
  \ee
 which is of the same form as (\ref{psieqm}),  instead of the free equation (\ref{freeeq}).
 A natural guess would be to add a term of the form
 \be
 \label{ges}
 -
	 \sum_{n=3}^\infty 
	 {{\hat\kappa^{n-2}} \over {n!}}
	 \{\wh   \Psi^n\}\, 
	\ee
	to the action. However, the independent fields in the theory are $\Psi,\wt\Psi$ and not $\Psi,\wh\Psi$, so the interaction should be a function of $\Psi,\wt\Psi$ and not $\Psi,\wh\Psi$. It will be seen in the next subsection that the simple guess (\ref{ges}), regarded as a function of $\Psi,\wt\Psi$, in fact does what is needed.  In the subsequent section background independence of the new action will be discussed. 
	
 \subsection{The New  Superstring Field Theory Action}

 The new SFT action for the string fields $\Psi,\wt\Psi$, is 
 \be\label{newact}
	S(\Psi,\wt\Psi)=- \tfrac{1}{2}  \langle \widetilde \Psi ,  
	Q \, {\cal G} \, \widetilde\Psi\rangle  
	+ \langle \widetilde \Psi ,   
	Q \, \Psi\rangle +\sum_{n=3}^\infty 
	 {{\kappa^{n-2}} \over {n!}}
	 \{\Psi^n\}\, -
	 \sum_{n=3}^\infty 
	 {{\hat\kappa^{n-2}} \over {n!}}
	 \{\wh   \Psi^n\}\, , 
	\ee
where
 \be
\wh    \Psi=  \Psi-{\cal G} \widetilde \Psi
  \ee
  There are two independent coupling constants $\kappa ,\hat \kappa$.
  Setting $\hat \kappa=0$ gives the Sen action (\ref{senact}) while setting both $\kappa ,\hat \kappa$ to zero gives the free action (\ref{free}).
  
  Varying the action with respect to 
  $\Psi$ gives the field equation
  \be
  \label{tifeeq}
  Q \widetilde  \Psi + \sum_{n=2}^\infty {{\kappa^{n-1}}\over n!} [ \Psi^n] - \sum_{n=2}^\infty 
	 {{\hat\kappa^{n-1}} \over {n!}}
	 [\wh   \Psi^n]=0\,
  \ee
  while varying with respect to $\widetilde  \Psi $ gives
  \be 
  Q \Psi - Q {\cal G} \widetilde \Psi + {\cal G}\sum_{n=2}^\infty 
	 {{\hat\kappa^{n-1}} \over {n!}}
	 [\wh   \Psi^n]=0\,
  \ee
  The second field equation can be rewritten as
   \be 
   \label{feqh}
  Q   \wh   \Psi + {\cal G}\sum_{n=2}^\infty 
	 {{\hat\kappa^{n-1}} \over {n!}}
	 [\wh   \Psi^n]=0\,
  \ee
  while acting on the first field equation with $ {\cal G}$ and adding to the second field equation gives
  \be
  \label{feq}
 Q \Psi +{\cal G}\sum_{n=2}^\infty {\kappa ^{n-1}\over n!} [ \Psi^n] =0 \,,
 \ee
 so that the field equations for $\Psi$ and $\wh \Psi$ are of the same form.
 In the NS-NS sector ${\cal G}=1$ and the general solution is given by finding
  solutions $\Psi$ and $\wh \Psi$ of the equations (\ref{feqh}),(\ref{feq}), and then $ \widetilde \Psi$ is given by
 $ \widetilde \Psi=  \Psi-\wh  \Psi
 $. Note that, as for Sen's SFT, the two field equations are completely decoupled: the field equation for $\Psi$ has no dependence on $\wh \Psi$ and vice versa, so that the solution for $\wh \Psi$ has no impact on the solution for $\Psi$.
  
  The action is invariant under the following gauge transformations:
  \be
	\label{gaugetrans2}
	\begin{split}
		\delta   |\widetilde\Psi\rangle  =& \  Q |\widetilde \Lambda\rangle  + \sum_{n=1}^\infty  {\kappa ^{n-1}\over n!}  \, [ \Lambda \Psi^n ]
		- \sum_{n=1}^\infty  {\hat\kappa ^{n-1}\over n!}  \, [(\Lambda- {\cal G}\widetilde \Lambda)\wh    \Psi^n ]
		\,, \\ 
		\delta  |\Psi\rangle  =& \  Q | \Lambda\rangle  + \sum_{n=1}^\infty  {\kappa ^{n-1}\over n!}  \, {\cal G} \,  [ \Lambda \Psi^n ]\,. 
	\end{split}
	\ee
The proof of gauge invariance uses the generalised Jacobi identities of the twisted $L_\infty$ algebra; the details will be given elsewhere \cite{HullPrep}.	
 It follows that  the transformation of  $\wh\Psi$ is
   \be
   \delta   |\wh\Psi\rangle  = \  Q |\wh \Lambda\rangle   
		+ {\cal G}\sum_{n=1}^\infty  {\hat\kappa ^{n-1}\over n!}  \, [\wh   \Lambda\wh    \Psi^n ]
\ee
where 
\be
\wh   \Lambda=\Lambda- {\cal G}\widetilde \Lambda
\ee
  which is of the same form as the transformation of $\Psi$. Thus the field equations   and gauge transformations   for $ \Psi$ and $\wh  \Psi$ are of exactly the same form. The gauge transformation of $\wt \Psi$ can then be written as
  \be
  \delta   |\widetilde\Psi\rangle  = \  Q |\widetilde \Lambda\rangle  + \sum_{n=1}^\infty  {\kappa ^{n-1}\over n!}  \, [ \Lambda \Psi^n ]
		- \sum_{n=1}^\infty  {\hat\kappa ^{n-1}\over n!}  \, [\wh   \Lambda\wh    \Psi^n ] \, .
\ee		
 Then  $\Psi$ transforms under the symmetries with parameter $\Lambda$,  $\wh \Psi$ transforms under the symmetries with parameter $\wh\Lambda$
  and $\wt\Psi$ transforms under both symmetries.
  
\section{Background Independence}. \label{backin}
 
\subsection{Background Independence in String Field Theory}
  
  There are some distinct but related notions of background independence in string field theory \cite{Sen:2017szq,Sen:1990hh,Sen:1993mh,Sen:1993kb,Sen:1994kx}.
 The simplest is an intrinsic property of a classical field theory and has been discussed in earlier sections for graviton actions.
 The action (\ref{gravactX}) for a free graviton $h$ in a background metric  $\bar g$  depends essentially on the background metric, while the non-linear action  (\ref{SEH}) 
 depends only on the total metric  $g=\bar g +\kappa h$ and so is  independent  of the way in which the total field is split into a background field and a fluctuation. As a result, it has the shift symmetry (\ref{shift}).

Similar remarks apply to the 2-graviton actions $S(\bar g, h, \hat h)$ considered earlier.
The quadratic action  (\ref{gravact2cov}) is background dependent.
The action (\ref{2gra1}) is non-polynomial in $h$ but quadratic in $\hat h$.
 The second term in the action (\ref{2gra1})  depends on both $\bar g$ and $\hat h$ and is background dependent. 
 The first term in the action  depends only on $g=\bar g +\kappa h$ and on its own would be background independent, but the full  action has no shift symmetry and  is background dependent.
 Sen's SFT, which led to the gravitational action (\ref{2gra1}), has similar properties. The action (\ref{senact}) is non-polynomial in $\Psi$ but is quadratic in $\wt \Psi$ and  the action is background dependent, as can be seen from the expansion of the action about a background solution in \cite{deLacroix:2017lif}.
 The field equation for $\Psi$ is 
background independent, however, while that for $\wt \Psi$ or $\wh \Psi$ is not.

The 2-graviton action (\ref{2grav})  is non-polynomial in both $h$ and $\hat h$ and can be written in terms of the total metrics $g=\bar g +\kappa h$, $\hat g=\bar g +\hat \kappa \hat h$ and so has the shift symmetry (\ref{shift2}). As  it depends only on the total metrics $g, \hat g$, it is background independent.
The new  SFT action (\ref{newact}), which led to the gravitational action (\ref{2grav}),  is non-polynomial in both $\Psi$ and $\wt \Psi$, with both sectors constructed about the same background. It will be seen in the next subsection that this theory is  independent  of the common background, with a shift symmetry so that the action depends only on total string fields.

This is to be contrasted with the notion of background independence discussed in \cite{Sen:2017szq,Sen:1990hh,Sen:1993mh,Sen:1993kb,Sen:1994kx}.
 Any  $c=15$ SCFT  constitutes a consistent background for classical superstring theory and  can be used to construct a superstring field theory,
 while any $c=26$ CFT  constitutes a consistent background for the bosonic string and can be used to construct a bosonic  string field theory.
 A longstanding question is whether different choices of (S)CFT can be regarded  as giving rise to the same string field theory. 
 Two (S)CFTs ${\cal C}$ and ${\cal C}'$ will give rise to  string field theories  ${\cal S}$ and ${\cal S}'$ respectively. 
 The first question is whether the string background defined by ${\cal C}'$ can be identified with a solution of the  string field theory  ${\cal S}$. 
 If it can, then there are two  string field theories for the background defined by ${\cal C}'$: there is ${\cal S}$ expanded about the solution and there is ${\cal S}'$. 
The key question is whether these are the same  string field theories,  related by a field redefinition.
For the bosonic string, this was analysed  in  \cite{Sen:1993mh,Sen:1993kb} for the case in which the  $c=26$ CFT ${\cal C}'$ is given by an infinitesimal marginal  deformation of the CFT ${\cal C}$ and it was shown that the two  string field theories are indeed equivalent. This  then implies that  any two CFTs related by a finite marginal deformation
also define the same  string field theory and so the  string field theory can be said to be locally independent of the background used to define it.\footnote{ In comparing the state spaces of different CFTs, it was necessary to introduce a connection on the space of CFTs \cite{Sen:2017szq,Sen:1993mh,Sen:1993kb,Ranganathan:1992nb,Ranganathan:1993vj}.}

In  \cite{Sen:2017szq}, the same questions were considered for  the superstring in the case of two $c=15$ SCFTs ${\cal C}$ and ${\cal C}'$  related by  an infinitesimal marginal NS-NS deformation.
The corresponding Sen  superstring field theories  ${\cal S}$ and ${\cal S}'$ were analysed and it was seen that the actions   for the two  string field theories were {\it not} the same. However, the field equations for $\Psi$ for the two  string field theories were shown to be related by a field redefinition and so it could be said that the field equation for the physical string field was background independent \cite{Sen:2017szq}. 

These two notions of background dependence are closely linked. For the bosonic string, the  string field theory is independent of the choice of classical background  solution and is also unchanged under marginal deformations of the  CFT used in its construction.
For the superstring, the Sen  superstring field theory action depends on the choice of classical background  and on the choice of SCFT,
while the field equation for $\Psi$ does not depend on either. The new action 
(\ref{newact}) is independent of the choice of classical background and the results of  \cite{Sen:2017szq}
suggest that the field equation for $\Psi$ should be unchanged (up to field redefinitions) under NS-NS marginal deformations of the defining SCFT.
It seems plausible that the arguments of \cite{Sen:2017szq,Sen:1990hh,Sen:1993mh,Sen:1993kb,Sen:1994kx}
  should extend to the statement that the action itself is unchanged (up to field redefinitions) under marginal deformations of the defining SCFT; this question will be addressed elsewhere.

\subsection{Background Independence for the New Superstring Field Theory Action}
  
 In this subsection we will set $\kappa=\hat \kappa=1$ for simplicity of presentation; the general case will be discussed in \cite{HullPrep}.
 Let
 \be
 {\cal F}(\Psi)= Q \Psi +{\cal G}\sum_{n=2}^\infty {1\over n!} [ \Psi^n]  
 \ee
so that the field equations (\ref{feqh}),(\ref{feq}) are
 \be
 {\cal F}(\Psi)= 0,
 \qquad
 {\cal F}(\wh   \Psi)= 0\, .
 \ee
Consider now expanding the string fields about a classical solution $(\overline { \Psi },\overline { \wt \Psi })$, so that
 \be
 { \Psi} =\overline \Psi +\psi , \qquad  
 \tilde {  \Psi} =\overline {\tilde \Psi }+\tilde \psi 
 \ee 
 giving rise to the corresponding expansion
 \be
 { \wh   {  \Psi} }={ \overline { \wh \Psi }}+\wh   \psi 
 \ee
 The background satisfies the classical field equations, so that
 \be
 {\cal F}(\overline\Psi)= 0,
 \qquad
 {\cal F}(\overline{\wh   \Psi})= 0
 \ee
 For a purely NS-NS solution these equations are sufficient, but for the other sectors it must further be imposed that
  $(\overline { \Psi },\overline { \wt \Psi })$ satisfy
 (\ref{tifeeq}).
 
 Then
  \be
 {\cal F}(\bar \Psi +\psi)= Q_{\bar \Psi } \psi +{\cal G}\sum_{n=2}^\infty {1\over n!} [ \psi^n]_{\bar \Psi }
 \ee
 where for any $A\in {\cal H}_c$
 \be
 Q_{\bar \Psi } A
  =QA+{\cal G}
 \sum_{n=1}^\infty {1\over n!} [{\bar \Psi }^n A]
 =QA+{\cal G}[{\bar \Psi }A]+\frac 1 2 {\cal G}[{\bar \Psi }^2A]+\dots
 \ee
 defines a map $ Q_{\bar \Psi }:  {\cal H}_c\to  {\cal H}_c$ 
 and for any $A_i \in {\cal H}_c$
 \be
 [A_1, A_2,\dots A_n]_{\bar \Psi }=
 [A_1, A_2,\dots A_ne^{\bar \Psi }], \qquad [A]_{\bar \Psi }=Q_{\bar \Psi } A
 \ee
  defines a map $  {\cal H}_c^n\to  \wt{\cal H}_c$. 
 The deformed BRST charge $Q_{\bar \Psi } $ and the  deformed brackets $[A_1, A_2,\dots A_n]_{\bar \Psi }$ satisfy the same
 algebraic conditions as $Q$ and $[A_1, A_2,\dots A_n]$ and so again define a twisted
  $L_\infty$  algebra.
  It will also be useful to define a deformed BRST operator acting on $ \wt {\cal H}_c$, with
  $ \wt Q_{\bar \Psi }:  \wt {\cal H}_c\to  \wt {\cal H}_c$ 
  defined by
   \be
 \wt Q_{\bar \Psi } \tilde  A 
  =Q\tilde A+ 
 \sum_{n=1}^\infty {1\over n!} [{\bar \Psi }^n ({\cal G}\tilde A)]
  \ee
  so that \cite{deLacroix:2017lif}
  \be
  {\cal G} \wt Q_{\bar \Psi }=Q_{\bar \Psi }{\cal G}\, .
  \ee
  
  Then   $ {\cal F}(\Psi)$ expanded around a solution gives
 \be
 {\cal F}(\bar \Psi +\psi)=
  {\cal F}_{\bar \Psi }(\psi)\ee
  where 
  \be
   {\cal F}_{\bar \Psi }(\psi)=Q_{\bar \Psi } \psi +{\cal G}\sum_{n=2}^\infty {1\over n!} [ \psi^n]_{\bar \Psi } \, .
 \ee
 The field equations $ {\cal F}(\Psi)= 0$ 
 become
  \be
    Q_{\bar \Psi } \psi +{\cal G}\sum_{n=2}^\infty {1\over n!} [ \psi^n]_{\bar \Psi }=0 \, ,
 \ee
which is  of the same form as (\ref{feq}) 
but with the BRST charge and brackets replaced by the deformed ones.
Similar relations hold for the field equation $ {\cal F}(\wh   \Psi)= 0$ but with all variables now hatted, so that
 \be
 {\cal F}({ \overline { \wh \Psi }} +\wh  \psi)=
  {\cal F}_{ \overline { \wh \Psi }}(\wh  \psi)\ee
  and can be expressed in terms of 
a deformed BRST charge $Q_{ \overline { \wh \Psi }}$ and  brackets $[A_1, A_2,\dots A_n]_{ \overline { \wh \Psi }}$.

Then the field equations expanded around a general solution involve two different BRST charges, $Q_{ \overline {   \Psi }}$ and $Q_{ \overline { \wh \Psi }}$, corresponding to expanding the two string fields about different backgrounds.
However, the original construction of the  string field theory involves a single SCFT,   and so uses 
the same BRST charge and $n$-brackets for all fields.
This corresponds to choosing backgrounds in which
\be
\label{peqbp}
{ \overline { \wh \Psi }}={ \overline {   \Psi }}
\ee
or equivalently
\be
{ \overline { \wt \Psi }}=0
\ee
Then if
 \be
 \label{ertw}
 {\cal F}(\overline\Psi)= 0
 \ee
 this gives a solution of the field equations derived from (\ref{newact}).
 For such solutions,  there is only one deformed BRST charge as $Q_{ \overline {   \Psi }}=Q_{ \overline { \wh \Psi }}$, and similarly there is only one set of brackets $[A_1, A_2,\dots A_n]_{ \overline {  \Psi }}=[A_1, A_2,\dots A_n]_{ \overline { \wh \Psi }}$.
 Here ${ \overline {   \Psi }}$ can be any solution of (\ref{ertw}) and need not be restricted to e.g.\ the NS-NS sector. 
Note that, in contrast, Sen's theory does not have non-trivial solutions   with ${   { \wt \Psi }}=0$.

The expansions are then
 \be
 { \Psi} =\overline \Psi +\psi , \qquad  
 \wt {  \Psi} = \tilde \psi , \qquad  
 { \wh   {  \Psi} }={ \overline {   \Psi }}+\wh   \psi 
 \ee 
  so that the field equations for $\psi,\hat \psi$ are 
  \be
   {\cal F}_{\bar \Psi }(\psi)=0,
   \qquad {\cal F}_{\bar \Psi }(\hat\psi)=0
   \ee
while the field equation (\ref{tifeeq}) for $\wt \Psi$ expanded around the background is
   \be
  \label{tifeeqex}
  \wt Q_{\bar \Psi } \widetilde  \psi + \sum_{n=2}^\infty {1 \over {n!}} [ \psi^n] _{\bar \Psi }- \sum_{n=2}^\infty 
	{1 \over {n!}} 
	 [\wh   \psi^n]_{\bar \Psi }=0\, .
  \ee
  
The  string field theory is constructed around a particular background using the BRST charge  $Q$ and $n$-brackets  $[...]$ for that background.
For another background  that is given by a solution $(\Psi,\wt\Psi)= ( \overline {   \Psi },0)$ of the original  string field theory defined by $Q$,  $[...]$,
there is a deformed BRST charge $Q_{\bar \Psi } $ and deformed $n$-brackets $[...]_{\bar \Psi }$.
Using these in
the new action (\ref{newact}) gives the action 
 \be\label{newact2}
	S_{\bar \Psi }(\psi,\wt\psi)=- \frac{1}{2}  \langle \widetilde \psi ,  
	Q_{\bar \Psi } \, {\cal G} \, \widetilde\psi\rangle  
	+ \langle \widetilde \psi ,   
	Q_{\bar \Psi } \, \psi\rangle +\sum_{n=3}^\infty 
	 { 1 \over {n!}}
	 \{\psi^n\}_{\bar \Psi } \, -
	 \sum_{n=3}^\infty 
	 {1 \over {n!}}
	 \{\wh   \psi^n\}_{\bar \Psi } \, , 
	\ee	
	for the new background.

The question now arises as to whether this action (\ref{newact2}) for the new background can be viewed as the {\it same}  string field theory as the original one	 constructed with $Q$,  $[...]$, i.e. whether the  string field theory can be viewed as being independent of the choice of background.	
Consider then the 	original action $S(\Psi,\wt\Psi)$ given in 	(\ref{newact}).
For the solution with $\wt\Psi=0$, 
\be 
S(\overline {   \Psi } ,0
)=0 \, .
\ee
The expansion of
 $S(\Psi,\wt\Psi)$  around the background with $ { \Psi} =\overline \Psi +\psi , \wt {  \Psi} = \tilde \psi$ gives, after a calculation,  precisely the deformed action (\ref{newact2}):
\be
S(\bar \Psi +\psi, \wt\psi)=S_{\bar \Psi }(\psi,\wt\psi) \, .
\ee
 In this sense, the new action is background independent and is a function of the total fields
$ { \Psi} =\overline \Psi +\psi , \wt {  \Psi} = \tilde \psi$. As a result, it has the shift symmetry
\be 
\label{SFTshift}
\delta\overline \Psi =A, \qquad \delta \psi =-A , \qquad   \delta\tilde \psi=0, \qquad \delta \hat \psi =-A
\ee

Note that this is {\it not} the case for Sen's action: the action constructed from the deformed BRST charge and brackets is distinct from the original one and so the action is not background independent in this sense and there is no shift symmetry.
Nonetheless, for Sen's theory with backgrounds with ${ \overline { \wt \Psi }}={ \overline {   \Psi }}$, the field equation  $ {\cal F}_{\bar \Psi }(\psi)=0$ for the   string field theory 
defined  for the background  solution using $Q_{\bar \Psi }$, $[\dots]_{\bar \Psi }$ 
is the same as the
the field equation for $\psi$ coming from expanding the original  string field theory using $Q $, $[\dots] $
expanded about the background with $\Psi=\bar \Psi +\psi$  as ${\cal F}(\bar \Psi +\psi)=
  {\cal F}_{\bar \Psi }(\psi)$, so that this  
  depends only on the total field $\Psi=\bar \Psi +\psi$.
   However, this is not the case for the   $\wh\Psi$ field equation in Sen's SFT as the two field equations $Q_{\bar \Psi }\wh\psi=0$
  and $Q\wh\psi=0$ are not related in this way, so that this field equation is background dependent.

\section{The Heterotic String}

The construction of Sen's heterotic string field theory is similar to that of the type II SFT, and again uses two  
	string fields, $\Psi$ and $\widetilde\Psi$, both annihilated by (\ref{Virg}).
	It is interesting that using a first-quantised approach to the heterotic string using the  bi-metric formulation of chiral scalars on the world-sheet also leads to a doubling of the heterotic string, with a physical  heterotic string and a  decoupled shadow one
	\cite{Hull:2025rxy}.
The string fields are states in the Hilbert space of  a $(1,0)$ SCFT with $c_L=15,c_R =26$, coupled to ghosts and bosonised superghosts.
The supersymmetric left-moving sector has picture number $-1$ for the   NS part of both string fields, picture number $-1/2$ for the R sector of the 
$\Psi$ field, and  picture number $-3/2$  for the    R sector of the 
$\widetilde \Psi$ field.
Then
	\be
	\label{HilbsHet}
	\begin{split} 
		& \Psi\in {\cal H}_c \equiv {\cal H}_{-1}\oplus {\cal H}_{-{1\over 2}}, \\[0.6ex] 
		& \widetilde\Psi\in \widetilde{\cal H}_c \equiv
		{\cal H}_{-1}\oplus {\cal H}_{-{3\over 2}}
			\, .
	\end{split}
	\ee
  The operator ${\cal G}:  \widetilde{\cal H}_c \to {\cal H}_c$  changes the picture number  and is defined by:
	\be
	\label{Gdefpc}
	{\cal G}\equiv \begin{cases} \hbox{${\bf 1} $   on ${\cal H}_{-1}$}\, ,\cr \hbox{${\cal X}_0$ on ${\cal H}_{-3/2}$}\, .
	\end{cases}
	\ee
Sen's action is again (\ref{senact}) and the new action for the heterotic string is again (\ref{newact}).
The field equations and symmetries are as before, but with the string fields now states in the Hilbert space of the $(1,0)$ SCFT.

\section{Discussion} \label{Discussion}

In this paper, a new superstring field theory action (\ref{newact}) was found by adding extra interactions to Sen's action so that both the physical superstring and the shadow superstring are interacting.

Some of the differences between these two SFTs can be seen by comparing the low energy effective actions for the gravitons.
Both theories have two gravitons $h,\hat h$ interacting with a background metric $\bar g$.
For Sen's theory, the low energy effective action for the graviton is (\ref{2gra1}) which is background-dependent and is an interacting theory for $h$ but a free theory for $\hat h$.
It has the background diffeomorphism symmetry (\ref{difff}) and the symmetries (\ref{exo1}),(\ref{exo2}) arising from the SFT gauge transformations.
The $\lambda$-transformations (\ref{exo1}) are a symmetry for any background while the $\hat \lambda$-transformations (\ref{exo2}) are a symmetry only for Ricci-flat backgrounds. The absence of a shift symmetry means that the $\lambda$ and $\hat \lambda$ symmetries are not directly relatable to diffeomorphisms.

For the new action, the low energy effective graviton action is (\ref{2grav}). It is background-independent and is an interacting theory for  both $h$ and  $\hat h$. It has  the  $\rho$ and $\hat \rho$ symmetries (\ref{dff1}),(\ref{dff2}) 
and the diagonal subgroup with $\rho=\hat \rho$ is
 the  diffeomorphism symmetry (\ref{rdif}).
 It also has the $\lambda$ and $\hat \lambda$ symmetries (\ref{lams}),(\ref{lamsh}) and it is these that  
 arise from the SFT gauge transformations.
The background independence means that it now also has the shift symmetry (\ref{shift2}). Then each transformation in
the diagonal subgroup of the
$\lambda$ and $\hat \lambda$ symmetries is a diffeomorphism combined with a shift symmetry.

The background-independent action (\ref{2grav}) can   be expanded about two
 different background metrics (\ref{genbac}) giving an action $S(\bar g, \bar {\hat g},h, \hat h)$ with two shift symmetries
(\ref{shift3}),(\ref{shift4}) as well as the $\rho$ and $\hat \rho$ symmetries (\ref{dff1}),(\ref{dff2}). 
The $\lambda $ symmetry (\ref{lams}) is then a $\rho$-diffeomorphism (\ref{dff1}) combined with a $\alpha$-shift (\ref{shift3}), while the $\hat\lambda $ symmetry (\ref{lamsh})  is a $\hat\rho$-diffeomorphism (\ref{dff2}) combined with a $\hat\alpha$-shift (\ref{shift4}).
In this way, some of the strange features of the \lq exotic diffeomorphisms' are understood.

The full set of massless fields consist of the  physical IIA or IIB supergravity multiplet together with a second supergravity multiplet in the shadow sector. Given the structure of the gravity effective action, the natural guess for the low energy effective action would be the usual supergravity action for the physical supermultiplet minus the same action for the shadow supermultiplet.
However, the structure of the theory is much more interesting.
It was to be suspected that the action couldn't be so simple as there isn't a conventional IIB supergravity action because of the presence of the self-dual gauge field.

For the fields in the NS-NS sector, the action is indeed the difference between the physical and shadow actions. However, in the RR sector the term $\langle \widetilde \Psi ,   
	Q \, \Psi\rangle$ in the action gives terms in the effective action of the form
	$$
	\int G\wedge d\tilde C$$
	where $G$ is a   $q$-form RR field strength arising from the string field $\Psi$ while $\tilde C$ is a  $q-1$-form RR gauge potential arising from $\widetilde \Psi$ \cite{Mamade:2025jbs}. 
	Such terms with $q=1,3,5,7,9$ arise  for the IIB string and those with $q=2,4,6,8,10$ occur for the IIA string. 
	The action has many interesting features  \cite{Hull:2025bqo}.  It   is democratic in that it treats each RR gauge field and its dual in the same way, and it successfully gives an action for the self-dual gauge fields when $q=5$. 
	It is acted on by both the $\rho$-diffeomorphisms and the $\hat\rho$-diffeomorphisms and leads to unusual couplings to both metrics $g, \hat g$ of the kind discussed in \cite{Hull:2023dgp}.
The physical $q$-form   field strength $F$ is constructed from the fields $G,\tilde C$ and couples only to the physical metric $g$, as a result of the unusual couplings, while the shadow  sector field strength $\tilde F$ does not couple to $g$.
The RR effective action will be discussed further elsewhere.

In the new SFT action presented here, both sectors are constructed about the same background. However, it is natural to ask whether it is possible to use a different background for each sector.
In the discussion of background independence of the new SFT action, we have focussed on background solutions of the form (\ref{peqbp}) and have set $\kappa=\hat \kappa$. This corresponds to constructing the full theory as an expansion of both sectors about a single SCFT. 
As has been seen, this theory has the important feature that it
depends just on the two total string fields, not serperately on the background and the fluctuations. However, these two total string fields can then each be expanded about a different background, with two different coupling constants, $\kappa,\hat \kappa$.
 These two backgrounds
correspond to two different SCFTs with different BRST charges and brackets. One is used in the field equation for $\psi$ and the other for the field equation for $\hat \psi$.
This is the analogue of the gravitational action (\ref{2grav}) expanded about two different background metrics (\ref{genbac}). 
This gives a more general SFT action based on two SCFTs and will be discussed in detail in \cite{HullPrep}.

As in \cite{Sen:2017szq,Sen:1990hh,Sen:1993mh,Sen:1993kb,Sen:1994kx}, the background independence discussed here is limited.
In  \cite{Sen:2017szq,Sen:1993mh,Sen:1993kb,Sen:1994kx}, the changes of background considered were between ones related by marginal deformations of the underlying SCFT.
In the string field theory defined by a particular SCFT, this seems to correspond to  changing to a background defined by a solution of that string field theory.
Thus the changes in background considered are between ones that are ``close", as would be the case in which the changes in background came from varying moduli. It remains a big challenge to go beyond this in string field theory, to formulate a theory that be used on any background, not just on a family of closely related backgrounds. The limited background independence achieved here and the accompanying understanding of the symmetries and the relation to gravity may be a first step in that direction.


\section*{Acknowledgments}

This research   was supported by   the STFC Consolidated Grant    ST/X000575/1.

\bibliographystyle{JHEP}
\bibliography{refs}

\end{document}

%% file: macros.tex
\usepackage{jheppub}
\usepackage{physics}
\usepackage{tensor}
\usepackage{amsmath}
\usepackage{amssymb}
\usepackage[colorlinks=true]{hyperref}
\usepackage{xcolor}
\usepackage{cancel}
\usepackage{dsfont}
\usepackage{tensor}
\usepackage{MnSymbol}
\usepackage{extarrows}

\usepackage{titlesec}
\makeatletter
\g@addto@macro\bfseries{\boldmath}
\makeatother

\usepackage{framed}
\usepackage[most]{tcolorbox}
\usepackage{xcolor}
\colorlet{shadecolor}{gray!10}

\hypersetup{
    linkcolor=blue
}

\newcommand{\wt}{\widetilde}
\newcommand{\wh}{\widehat}
\newcommand{\be}{\begin{equation}}   
	\newcommand{\ee}{\end{equation}}

\usepackage{tikz}

\usetikzlibrary{3d}
\usetikzlibrary{calc, decorations.markings}

\pgfkeys{/tikz/.cd, view angle/.initial=0, view angle/.store in=\picangle}

\tikzset{
    horizontal/.style={y={(0,sin(\picangle))}},
    vertical at/.style={x={([horizontal] #1:1)}, y={(0,cos(\picangle)cm)}},
    every label/.style={font=\tiny, inner sep=1pt},
    shorten/.style={shorten <=#1, shorten >=#1},
    shorten/.default=3pt,
    ->-/.style={decoration={markings, mark=at position #1 with {\arrow{>}}}, postaction={decorate}},
    ->-/.default=0.5,
    dark plane/.style={rotate around x=\angledp,canvas is xy plane at z=0,draw=cyan,fill=cyan!25},
}